\documentclass[12pt,preprint]{aastex}

\def\lim{{\rm lim}}

\def\bol{{\rm bol}}
\begin{document}

\title{Sensitivity of Transit Searches to Habitable Planets}

\author{Andrew Gould, Joshua Pepper, D.\ L.\ DePoy}
\affil{Department of Astronomy, The Ohio State University,
140 W.\ 18th Ave., Columbus, OH 43210}
\email{gould,pepper,depoy@astronomy.ohio-state.edu}

\singlespace

\begin{abstract}
Photon-limited transit surveys in $V$ band are in principle about 20 times 
more sensitive to planets of fixed size in the habitable zone around M stars
than G stars. In $I$ band the ratio is about 400.  The advantages that
the habitable zone lies closer and that the stars are smaller (together 
with the numerical superiority of M stars)  more
than compensate for the reduced signal due to the lower luminosity of
the later-type stars.  That is, M stars can yield reliable transit 
detections at much fainter apparent magnitudes than G stars.
However, to achieve this greater sensitivity, the later-type
stars must be monitored to these correspondingly fainter magnitudes, which
can engender several practical problems.  We show that with modest
modifications, the {\it Kepler} mission could extend its effective
sensitivity from its current $M_V=6$ to $M_V=9$.  This would not capture
the whole M dwarf peak, but would roughly
triple its sensitivity to Earth-like planets in the habitable zone.  
However, to take
advantage of the huge bump in the sensitivity function at $M_V=12$
would require major changes.

\end{abstract}
\clearpage
 
\section{Introduction
\label{sec:intro}}

	The one confirmed transiting planet discovered to date, 
HD209458b \citep{char00}, lies only $10\,R_\odot$ from its host
G star, and therefore well inside the so-called ``habitable zone'',
where water could exist in its liquid state.  All ongoing transit 
surveys are primarily sensitive to such ``hot Jupiters'' because
their large diameters give rise to relatively strong $(\sim 1\%)$
signals, while their proximity to their host increases the probability
of transits occurring \citep{how00,brown99,mal02,udal02,
burke02,str00,str02}.

	While gas giants are not expected to themselves be habitable
regardless of their location, they could have water-laden moons like
Europa, Ganymede, and Callisto, which could support life if the
planet lay in the habitable zone.  Future ground-based surveys could
be sensitive to such habitable-zone giant planets.  
Moreover, the planned {\it Kepler} 
mission\footnote{http://www.kepler.arc.nasa.gov/} has as its primary
goal the detection of planets down to Earth size in or near the habitable
zone.  The {\it Kepler} targets would be even more direct analogs of the Earth.

	Here we investigate the sensitivity of transit surveys to
planets in the habitable zone as a function of stellar type.  We
derive two remarkable conclusions.  First, transit surveys are most
sensitive to habitable planets orbiting middle M stars
($M_V\sim 12$), despite the fact that they are most sensitive overall 
to planets orbiting early G stars ($M_V\sim 4$).  Second, to achieve
this sensitivity to dim stars, the survey's magnitude limit must be
an extremely strong function of spectral type.  Specifically, it must
be about 7 magnitudes fainter for middle M stars than G stars.
Since such variable requirements can significantly affect survey design,
they deserve careful analysis and consideration.  For example, we show
that unless the {\it Kepler} mission extends its magnitude limit for M stars
well beyond its current (type independent) limit of
$V= 14$, it will lose most of its potential sensitivity to habitable planets.

\section{Sensitivity to Habitable Planets
\label{sec:sensitivity}}

\citet{pgd} showed that if every star of luminosity $L$ and radius $R$
has a planet of radius $r$ in a circular orbit of semi-major axis $a$, 
then a photon-noise limited transit survey will detect
\begin{equation}
N(L,R,r,a) = {\Omega d_0^3\over 3}n(L,R)\eta
{R_0\over a_0}
\biggl({L \over L_0}\biggr)^{3/2}
\biggl({R \over R_0}\biggr)^{-7/2}
\biggl({a \over a_0}\biggr)^{-5/2}
\biggl({r \over r_0}\biggr)^{6}
\label{eqn:ndetect}
\end{equation}
planets.  Here $n(L,R)$ is the number density of stars of the specified
type, $\Omega$ is the angular area of the survey, and $d_0$ is the
distance out to which an equatorial transit can just be 
reliably detected for the fiducial parameters $L_0$, $R_0$, $r_0$, and
$a_0$.  The numerical factor $\eta\simeq 0.719$ arises because the
volume probed by the survey is smaller by a factor $y^{3/2}$
for non-equatorial transits, where $y$ is the ratio of the transit
chord to the stellar diameter.  For the most part we will be comparing the 
detectability
of planets of the same size around different stars.  Hence, except when
we are evaluating the normalization of equation (\ref{eqn:ndetect})
for a specific example, we will ignore the last factor.

The equilibrium 
temperature of a planet (and thus whether or not water can exist
in liquid phase on the planet surface) is the same if
 $a\propto L_\bol^{1/2}$, 
where the
bolometric luminosity $L_\bol$ is to be distinguished from $L$, the
luminosity in the band of observation.  Hence, the relative sensitivity
of a transit survey to habitable planets is
\begin{equation}
N \propto n L^{3/2} R^{-7/2} L_\bol^{-5/4}.
\label{eqn:ndetprop}
\end{equation}
Here we assume that all planets
have the same albedo and neglect atmospheric effects.
If we compare, for example, G stars ($M_V=5$) with middle
M stars ($M_V=12$), the ratios of the various factors are
$N_{12}/N_{5} \sim 6\times 630^{-3/2}\times 4^{7/2}\times 80^{5/4}\sim 10$.
Note that if we were comparing detectability at the {\it same
semi-major axis} rather than the {\it same habitability}, the last
factor would not have entered, and the ratio would have been 0.05.
Hence, while the sensitivity to planets in general is completely
dominated by G stars, the sensitivity to habitable planets is completely
dominated by M stars.  That is, the lower luminosities (and so smaller
semi-major axes) combined with the smaller radii and greater numerical
density of M stars more than compensate for the reduced photon counts.

	In Figure \ref{fig:one}, we show the sensitivity to habitable
``Earths'' and to Earths all at the same semi-major axis (taken to be 1 AU).
That is, the histograms show the total number of planets $N$ that will be
detected as a function of $M_V$ assuming that each star in the
field has one Earth-size planet in the habitable zone or, respectively,
one such planet at 1 AU.  (The two histograms cross at $M_V=5$ because
the habitable zone is then at 1 AU).
The absolute normalization of this plot is set according to the 
characteristics of the {\it Kepler} mission
($7.8\times 10^8\,e^-\,\rm hr^{-1}$ at $V=12$, $\Omega=105\,\rm deg^2$,
$A_V=0.3$, mission-total S/N=8 required for detection), but the
form of the histogram would be the same for any photon-limited survey.
The normalization for any other planet size should be multiplied by
a factor $(r/r_\oplus)^6$, and the normalization for any other
fixed semi-major axis should be multiplied by $(a/{\rm AU})^{-5/2}$ 
(see eq.\ [\ref{eqn:ndetect}]).  The figure
is constructed assuming that detection is in $V$ band.  
The effect of substituting other bands is approximately to change the
slope of the histogram.  For example, since the slope of the main
sequence is $d M_V/d(V-I) =3.37$ \citep{reid91}, substitution of $I$
band would lead to an increase of slope 
$d\Delta \log N/d M_V = (3/2)\times 0.4/3.37 = 0.178$.  That is, middle M
stars would gain relative to G stars by an additional factor of
$10^{7\times 0.178} = 18$.

	To compute these histograms, we follow the procedure of \citet{pgd}.
An important feature of the color-magnitude diagram is that while
the main sequence is fairly narrow for $M_V>6$, it broadens for brighter stars
(due to faster stellar evolution),
so that a star of a given $M_V$ can have a large range of colors.
Thus, for the well-defined lower main sequence, $M_V>6$, we
consider the luminosity function \citep{bessell93,zheng01}
in 1 mag bins, and evaluate the stellar radius at the center of each bin
using the linear color-magnitude relation of \citet{reid91}, the
color/surface-brightness relation of \citet{vanb}, and the $VIK$
color-color relations from \citet{bessell88}.  

	On the other hand, for the upper main sequence $M_V<6$ we evaluate 
the histograms directly using the
Hipparcos catalog \citep{hip}.  For example, the luminosity function for
$M_V=4$ is computed by summing $\sum_i [(4/3)\pi D_i^3]^{-1}$ over all stars
within the Hipparcos completeness limit, $V<7.3$, having $3.5<M_V<4.5$,
and lying within 50 pc.  The distance $D_i$ is the minimum of 50 pc
and the distance at which the star would have $V=7.3$.  Then, the
constant-semi-major-axis histogram is computed by summing (and
appropriately normalizing) 
$\sum_i L_i^{3/2}R_i^{-7/2}[(4/3)\pi D_i^3]^{-1}$,
while the habitable-zone histogram is found by summing
$\sum_i L_i^{3/2}R_i^{-7/2}L_{\bol,i}^{-5/4}[(4/3)\pi D_i^3]^{-1}$.
The stellar radii are determined from Hipparcos/Tycho $(B_T,V_T)$ photometry
and the color/surface-brightness relation of \citet{gm}, ultimately
derived from \citet{vanb}.
We evaluate $L_\bol$ using bolometric corrections as a
function of $V-I$ color derived from \citet{bm} and \citet{bessell88} at the
bright end and \citet{rg} at the faint end.

\section{Magnitude Limits for Habitable Planets
\label{sec:maglims}}

	To achieve the sensitivities calculated in \S~\ref{sec:sensitivity}, 
one must analyze the light curves of all the stars being probed.  
This statement would appear so obvious as not to be worth mentioning.  However,
as we now show, dim stars are being probed at substantially fainter
apparent magnitudes than are their more luminous cousins.  Hence, to
avoid losing most of the sensitivity of a transit experiment
one must set different magnitude limits for stars of different $M_V$.

Let $m_\lim$ be the apparent magnitude at which the survey achieves the 
minimum acceptable signal-to-noise ratio (S/N) for a planet of radius $r$
and semi-major axis $a$ circling an $M_V=5$ star.  Consider an identical
planet circling an $M_V=12$ star in the same orbit and at the same apparent
magnitude.  Since the star's radius is
a factor 4 smaller, the S/N will be a factor $4^{2}\times 4^{-1/2}=4^{3/2}$
times larger.  The first factor comes from the fact that the planet occults
a larger fraction of the stellar surface, the second from the reduced 
total time spent in transit (``duty cycle'') 
due to the smaller radius.  Now, the bolometric luminosity of
the dimmer star is down by a factor 80, so to keep in the habitable zone,
the planet must be moved closer by a factor $80^{1/2}$.  This increases the
transit duty cycle by the same factor and so increases the S/N by $80^{1/4}$.
Combining these two effects with the usual dependence of S/N on flux implies
S/N $\propto {\rm flux}^{1/2}L_\bol^{-1/4}R^{-3/2}$.  Hence, to maintain the 
same S/N, one must increase the magnitude limit by,
\begin{equation}
\Delta m_\lim = 0.5\Delta M_\bol - 7.5\Delta \log R.
\label{eqn:deltavlim}
\end{equation}
For the example just given,
$\Delta m_\lim = 6.9$, i.e., a factor 570 in flux.  Note that
$\Delta m_\lim$ does not depend on the passband of observation.  The dashed
curve in Figure \ref{fig:one} shows the difference 
in the maximum distance modulus relative to the Sun 
at which a habitable planet (of given radius) can be detected.  Specifically, 
$\Delta(m-M)_0 = (V_\lim - M_V) - (V_{\lim,\odot} - M_{V,\odot})$.
For $V$ band observations, this offset is roughly
constant over 10 magnitudes.  That is, the limiting magnitude required
to achieve the sensitivities shown by the bold histogram increases 
approximately
in lock step with $M_V$.  This behavior can have major consequences for the
design of transit experiments.

\section{Discussion
\label{sec:discuss}}

\begin{figure}
\plotone{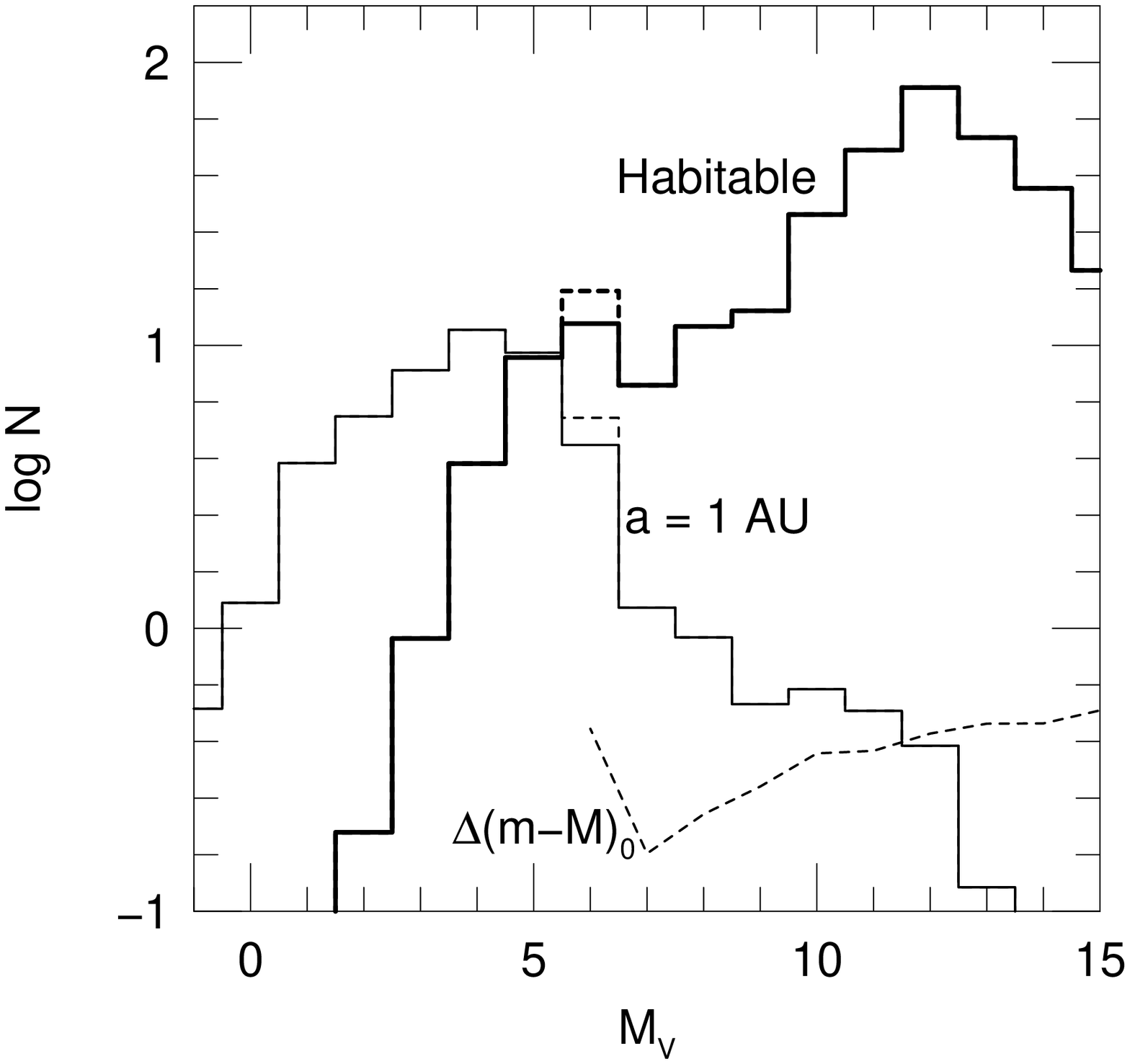}
\caption{\label{fig:one}
Total number of planets detected in the habitable zone ({\it bold histogram}) 
and in 1 AU orbits ({\it solid histogram}) as functions of
absolute magnitude $M_V$ (assuming every star has one such planet).
The absolute normalizations have been set to Earth-size planets and to 
the characteristics of the {\it Kepler} mission, but the form
is completely general assuming observations in $V$ band.  In $I$ band,
the slope would be tilted upward toward faint stars.  Extinction has
been taken into account, but stellar variability and binarity have not.
The calculation uses different methods for early ($M_V\leq 6$) and late
($M_V\geq 6$) stars.  The solid and dashed histograms compare the two
methods in the one bin of overlap.  The {\it dashed curve} shows the
difference (relative to Sun-type stars) in the limiting distance modulus 
to which stars of different $M_V$ must be monitored to achieve the sensitivity
shown by the ``habitable zone'' histogram.  Since this curve is basically
flat, dimmer stars must be observed to the same distance and hence to much
fainter apparent magnitudes.
}\end{figure}

Photon-limited observations are not routinely achievable over very large
dynamic ranges.  A number of effects can intervene at fainter magnitudes.
First, once the stars fall below the sky, the number of detected
systems no longer falls as $L^{3/2}$ (see eq.\ [\ref{eqn:ndetect}])
but as $L^3$.  That is, once the sky is reached, detections would
fall off by a factor $\sim 4$ per magnitude relative to the bold histogram
in Figure \ref{fig:one}.  Second, if the exposure times are set so as
not to saturate the brightest target stars (say 2 mag brighter than the
mag limit appropriate for $M_V=5$), then the flux levels will be
a factor $\sim 4000$ times lower at the mag limit appropriate for $M_V=12$.
For many observing setups, this would render the flux levels 
comparable to the read noise.
Third, a field that is very uncrowded at one magnitude may well be extremely
crowded 7 magnitudes fainter, and this may interfere with doing
photon-limited photometry.  Hence, it is unlikely that all of the
potential shown in Figure \ref{fig:one} can be achieved in any given
practical experiment.  Nevertheless, this potential is so great that it
is worth thinking about how to achieve as much as possible.

\begin{figure}
\plotone{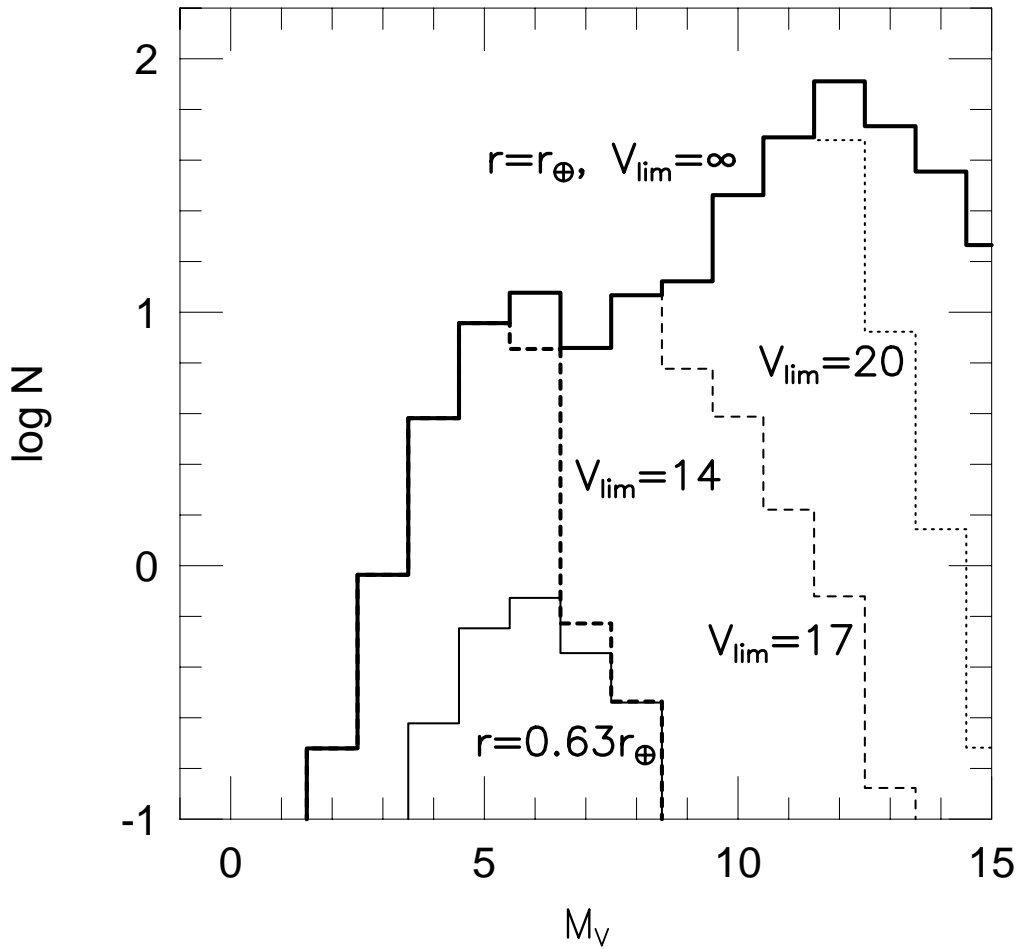}
\caption{\label{fig:two}
Total number of planets detected normalized to characteristics of {\it Kepler}
and under various assumptions.  
{\it Bold solid histogram}: each star has one Earth-size planet in the 
habitable zone and all stars are monitored regardless of magntiude (same
as Fig.\ \ref{fig:one}).
{\it Bold dashed histogram}: same planet distribution, but now only stars
$V<14$ are monitored.
{\it Thin dashed histogram}: same planet distribution, 
$V<17$ stars are monitored.
{\it Dotted histogram}: same planet distribution, 
$V<20$ stars are monitored.
{\it Thin solid histogram}: planets have radius $r=0.63\,r_\oplus$ and only
$V<14$ stars are monitored.
}\end{figure}

We illustrate the type of trade-offs involved by considering the
{\it Kepler} mission, the only transit experiment proposed to date
whose primary target is planets in the habitable zone.  
The {\it Kepler} design calls for recording only stars $V<14$, presumably
in order to minimize data transmission.  The mag limit required to detect 
Earth-like planets around Sun-like stars at S/N=8
is $V_{\lim,\odot}=13.6$.  Hence, from the dashed curve in 
Figure \ref{fig:one},
the required mag limit at $M_V=6$ is 
$V_\lim=V_{\lim,\odot} - M_{V,\odot} + M_V + \Delta(m-M)_0 = 14.4$, 
i.e., already somewhat fainter than
the $V=14$ limit adopted by {\it Kepler}.  The volume probed, and hence the
detections,
fall relative to the bold histogram by a factor $(10^{0.2})^3\sim 4$ 
for each magnitude beyond $M_V=6$.  See Figure \ref{fig:two}.
Hence, assuming that every star has one planet in
the habitable zone, only a total of 20 will be 
detected\footnote{Actually, we have not taken account of stellar
variability in this analysis, nor of binaries.  Each of these might
plausibly reduce detections by 25\%, the former by making transits
unrecognizable, the later by making habitable planets dynamically
unstable and by increasing the total light in the aperture.  Hence,
the true detection rate for this optimistic scenario should be reduced
by roughly 50\%.}.

Keeping the photometry information for stars to $V=17$, for example, 
should increase the number of habitable-planet detections
by almost a factor 3, 
while keeping stars with $V<20$ would increase this
number by a factor 9. See Figure~\ref{fig:two}.  Of course, 
there are many obstacles to 
keeping information on stars this faint.

For example, there would be an enormous number of giant stars contaminating 
such a deep sample.
In fact, however, 
distinguishing the later-type dwarfs from the much more numerous
giants of similar colors is straightforward.  As shown by \citet{gm}, 
they can easily be identified on a reduced proper motion (RPM) diagram
constructed using data from USNO-A \citep{usnoa1,usnoa2} and
2MASS \citep{2mass}.  With USNO-B \citep{usnob} it should be possible
to extend coverage to fainter magnitudes and to achieve higher precision
as well.  The total number of such stars $M_V\leq 12$ is only 
$\sim 3\times 10^4$, far fewer than the $\sim 10^5$ giants in the field
that could be eliminated using the same RPM diagram \citep{gm}.

A more difficult problem is crowding.  The {\it Kepler} point spread function
(PSF) is deliberately defocused to $10''$, meaning that the crowding limit
is about $10\,\rm arcmin^{-2}$.  At the adopted {\it Kepler} sight line,
$(l,b)=(69.6,5.7)$, this limit is reached at $R_{\rm USNO}\sim 18.2$
(as determined by a query of the USNO-A catalog).
If {\it Kepler} were redirected to $(l,b)=(70,15)$, the density of
stars down to $R_{\rm USNO}=18.2$ would be $3.5\,\rm arcmin^{-2}$,
approximately 3 times lower.  Hence, recovery of faint stars would be
much easier.  Presumably, {\it Kepler} has chosen to look right in
the Galactic plane because of the higher overall star density.  However,
since the stars useful for transits mostly lie within 500 pc, a field
at $b=15^\circ$ would lie only 130 pc above the plane, where the
density of such target stars is only slightly lower than at the plane.
Hence, the crowding problem could be substantially ameliorated at small
cost.

The last problem is sky.  At high ecliptic latitude, the sky in space
is $V\sim 23.3\,\rm arcsec^{-2}$, or $V\sim 17$ in a $10''$ PSF.  Hence,
{\it Kepler} completeness appears to be fundamentally limited to stars $M_V<9$.
As shown in Figure~\ref{fig:two}, detections from stars at $M_V\geq 10$ 
would be highly suppressed.  Only contracting the PSF (which has been
deliberately defocused to improve the photometry) could overcome this
difficulty.  (Note that at $V\sim 17$, crowding would not be a serious
problem even for the current Galactic-plane line of sight.)

While the above remarks apply specifically to {\it Kepler}, any transit
experiment attempting to detect habitable planets would face similar
constraints and trade-offs.

Finally, we note that the results reported here are a strong function of
planet size.  As remarked in \S~\ref{sec:sensitivity}, the absolute
normalizations in Figure \ref{fig:one} scale as $r^6$.  By an argument
similar to the one given in \S~\ref{sec:maglims}, 
$\Delta m_\lim = 10\Delta \log r$.  For example, for $r=0.63\,r_\oplus$,
the normalizations in Figure \ref{fig:one} would be reduced by a factor 16,
while the required $V_\lim$ at each $M_V$ would be brighter by 2 mag.
Hence, in {\it Kepler}'s current configuration, it would retain full 
sensitivity to $M_V\sim 8$ stars, and so would be able to detect $\sim 3$ 
planets (see Fig.~\ref{fig:two}), 
while if its coverage were extended 3 mag to $V_\lim\sim 17$, it could
probe the M-dwarf peak, and so detect $\sim 8$ planets.


\acknowledgments
We thank Scott Gaudi for his detailed comments on the manuscript. 
This work was supported by grant AST 02-01266 from the NSF.


\clearpage

\end{document}